\documentstyle[12pt,twoside,fleqn,espcrc1,epsfig]{article}

\newcommand\Jou[4]{{#1} {\bf #2} (#4) #3}

\newcommand\NPA{{Nucl. Phys.} A}
\newcommand\NPB{{Nucl. Phys.} B}

\newcommand\PRL{Phys. Rev. Lett.}
\newcommand\PRC{{Phys. Rev.} C}
\newcommand\PRD{{Phys. Rev.} D}

\newcommand\p{\pi}
\newcommand\q{\theta}

\renewcommand\t{\tau}

\newcommand\co{{\cal O}}

\newcommand\dd{\mbox{d}}

\newcommand\ra{\rightarrow}

\newcommand\mrm[1]{{\mathrm{#1}}}

\newcommand\be{\begin{equation}}
\newcommand\ee{\end{equation}}
\newcommand\bea{\begin{eqnarray}}
\newcommand\eea{\end{eqnarray}}
\newcommand\ba{\begin{array}}
\newcommand\ea{\end{array}}

\newcommand\ederef[2]{Eqs.~(\ref{#1}) and (\ref{#2})}

\newcommand\fref[1]{Fig.~\ref{#1}}

\newcommand\bfi{\begin{figure}}
\newcommand\efi{\end{figure}}
\newcommand\bpi[1]{\begin{picture}#1}
\newcommand\epi{\end{picture}}

\newcommand{\ncom}{\newcommand}
\ncom{\lan}{\langle}
\ncom{\ran}{\rangle}
\ncom\fx{\!\!\!\!}
\ncom\lqcd{\Lambda_{\mathrm{QCD} }}

\title{Condensation of a Strongly Interacting Parton Plasma into
a Hadron Gas in High Energy Nuclear Collisions}

\author{S.M.H. Wong
   \address{School of Physics and Astronomy, 
            University of Minnesota,  
            Minneapolis, MN 55455, U.S.A.}
        \thanks{The author is grateful to the organizing committee
                for a fellowship to cover the local conference expenses.
                This work was supported by the U.S. Department of Energy under
                grant DE-FG02-87ER40328.} 
}

\begin{document}

\maketitle

\begin{flushright}
\vspace{-5.0cm}
\footnotesize \sffamily NUC-MINN-99/10-T
\end{flushright}

\vspace{3.50cm}
\abstract{
We examine the effects of color screening on the transition of a parton 
plasma into a hadron gas at RHIC energies. It is found that as expected, 
color screening posed itself as a significant barrier for hadronization. 
Parton-hadron conversion would therefore be delayed and prolonged when 
compared to that occuring in a vacuum. Due to the on-going expansion, the 
resulting hadron densities are lowered. Parton equilibration is also shown 
to be seriously disrupted in the process. 
}

\section{Introduction}
\label{sec:intro}

Heavy ion collision experiments at the Relativistic Heavy Ion Collider
(RHIC) at Brookhaven are going to study the possibility of the 
recreation of a deconfined parton plasma. This plasma consists of gluons,
quarks and antiquarks which can roam freely in a region of sizable
spatial extent. This is made possible by the high energies of the
partons through the asymptotic freedom property of QCD and
the color screening property of a dense medium of color charges. 
The latter weakens the otherwise very strong color fields responsible
for keeping the partons well hidden inside hadrons under normal
circumstances. This same screening property has the additional
advantage of removing the infrared divergence in the parton interactions
and thus there is no need to introduce an ad hoc and somewhat arbitrary
momentum cutoff as in the usual studies within perturbative QCD in a
vacuum setting. This signals important differences of a dense parton
system in comparison to an extremely dilute system as found in deep 
inelastic scattering or in $e^+ e^-$ annihilation. Because the latters 
have been studied for well over a decade, it is tempting and certainly 
easiest to apply all the acquired knowledge or understanding in exactly
the same way as done in these studies to heavy ion collisions. 
In view of the above mentioned differences in the environment, this
cannot be entirely correct. It is clear that one has to bear in mind 
these differences when building numerical models.

The time evolution of a parton plasma has been studied in various models
\cite{pcm1,pcm2,biro,wong1,wong2}. What we have learned are that gluon 
equilibration tends to be fast in both thermal and chemical aspects but this 
is not so for quarks and antiquarks. But regardless of the state of the
system, that is whether it is fully equilibrated or not, hadronization
will take place once the conditions are right. The medium effects such as the 
Landau-Pomeranchuk-Migdal effects of gluon emission and the shielding of 
infrared divergence have been considered in \cite{biro,wong1,wong2} and in 
\cite{pcm1,pcm2} only the former has been incorporated. All were done, however,
within the deconfined phase. In this talk, we are concerned with the next 
stage in the time evolution of the parton system, that is the transition 
into hadrons and the medium effects on that. In \cite{pcm3} a hadronization 
scheme was introduced at the end of the time evolution of the parton cascades 
but the role of the medium on hadronization, which we will consider here,
has been totally neglected. 

In our previous investigation, the interaction strength in a parton 
plasma was found to increase in time due to the decreasing average energy of the
system \cite{wong2}, this has the consequence of an increasing screening mass 
$m_D$ with time at least at RHIC for the duration that we have
investigated. The screening length therefore behaved in the opposite manner. 
We got a screening length $l_D \sim 0.4$ fm at the end of our time evolution
when the temperature estimates fell to $200$ MeV. This is unfavorable
for hadronization because this $l_D$ value is comparable to the typical common
hadron size. This fact is further reinforced by the lattice calculation of 
$m_D$ up to $\co(g^3)$ in \cite{kaj}. They found that it was larger than the 
leading order result with $m_D \sim 3.3 \;m_D^\mrm{LO}$. Using their results 
and choosing $\bar \lqcd \simeq 234$ MeV, at $T \sim 200$ MeV 
\be  m_D^\mrm{LO} \sim 405 \;\mrm{MeV} \; \Longrightarrow \;
     l_D \sim l_D^\mrm{LO}/3.3  \sim 0.16 \;\mrm{fm}         \nonumber 
\ee
and at $T \sim 150$ MeV
\be  m_D^\mrm{LO} \sim 332 \;\mrm{MeV} \; \Longrightarrow \;
     l_D \sim l_D^\mrm{LO}/3.3  \sim 0.20 \;\mrm{fm}    \; . \nonumber 
\ee
These are small sizes compared to most hadrons. So it is clear that
the color screening barrier to hadronization is not small in a parton 
plasma found at RHIC.

\section{Time Evolution Equations for the Parton-Hadron Conversion}
\label{sec:t-eqn}

In \cite{wong1}, we wrote down the equations for the time evolution of a
parton plasma undergoing one-dimensional expansion. The main ingredient
of our scheme is to combine the reduced Boltzmann equation, the relaxation
time approximation for the parton collision terms $C_i^{\mrm{p}}$ and 
their explicit perturbative construction. The resulting equations are
\be {{\dd f_i^{\mrm{p}}} \over {\dd \t}} \, \Big |_{p_z \t =\mrm{const.}} 
    \; = - \; { {f_i^{\mrm{p}}-f^{\mrm{p}}_{i\; eq}}
            \over {\q_i^{\mrm{p}}} }
    \; =   \; C_i^{\mrm{p}}      \; .
\label{eq:old_p_ev_eq}
\ee
With this combination, the distributions $f^\mrm{p}_i$ which describe
completely the time development of the system can be solved. 

To extend the time evolution beyond the parton phase and to try to learn
something about the medium effects on the parton-hadron conversion, it
is not necessary to have the full three-dimensional expansion. There is also
no need for a full set of hadrons. So we will continue with one-dimensional
expansion and only consider pions and kaons. Moreover these hadrons or resonances 
will be assumed to consist of only a quark and an antiquark $q\bar q'$ and thus
their formation will be from the clustering of $q$ and $\bar q'$. 
Since the parton plasma is gluon dominated, the gluons must be converted somehow 
into $q$ and $\bar q$. Perturbative conversion is highly inefficient
so a non-perturbative mechanism must be introduced. In \cite{fw,web}, exactly
such a gluon splitting mechanism was introduced for this very purpose
in the context of $e^+ e^-$ annihilations. Although our parton plasma is 
different from a parton shower, a term of similar nature 
$C^{\mrm{p}}_{i\;g\ra q}$ will be introduced in the time evolution equations. 
Together with some new confining terms $C^{\mrm{p}}_{i\; p\ra h}$ which describe 
the clustering of color singlet $q\bar q'$ pairs into resonances and the 
subsequent decay into hadrons, the time evolution equations become
\be {{\dd f_i^{\mrm{p}}} \over {\dd \t}} \, \Big |_{p_z \t =\mrm{const.}} 
    \; = - \; { {f_i^{\mrm{p}}-f^{\mrm{p}}_{i\; eq}}
            \over {\q_i^{\mrm{p}}} }
    \; =   \; C_i^{\mrm{p}} + C^{\mrm{p}}_{i\; g\ra q} 
            + C^{\mrm{p}}_{i\; p\ra h} \; .
\label{eq:new_p_ev}
\ee
Because of the parton-hadron conversion, an equation for each hadron
will also have to be introduced. Using the same method, we write
\be {{\dd f_i^{\mrm{h}}} \over {\dd \t}} \, \Big |_{p_z \t =\mrm{const.}} 
    \; = - \; { {f_i^{\mrm{h}}-f^{\mrm{h}}_{i\; eq}}
            \over {\q_i^{\mrm{h}}} }
    \; =   \; C^{\mrm{h}}_{i\;p \ra h} \; .
\label{eq:new_h_ev}
\ee
Because they are non-essential to our investigation so no hadron-hadron 
or parton-hadron interactions are present in \ederef{eq:new_p_ev}{eq:new_h_ev}. 
These form our basic set of equations for the conversion of a parton into a hadron 
gas. The explicit forms of the $C^\mrm{p}_i$'s and $C^\mrm{h}_i$'s can be found 
in \cite{wong3}. We stress that no medium effect on hadronization has yet 
been included.

To understand how medium effects such as color screening would affect
hadronization, we use the following physical picture. Each hadron
must have a certain physical size which can be thought of as the internal
separation $b$ of the $q$ and $\bar q'$ pair. Because of the internal motion,
this separation is not fixed and smaller separations should be more
favorable. So this likelihood can be parametrized by a distribution
$F(b)$. Therefore for a hadron or resonance existing inside a color 
screening medium, there is a chance that it will dissolve depending on
whether $b<l_D$ or $b>l_D$ which we represent by $P_<$ and $P_>$, respectively.
They are related to the distribution by 
$     P_< = \int^{l_D}_0 \dd b \; F(b)$ 
and
$     P_> = \int^\infty_{l_D} \dd b \; F(b)$. 
They also dictate whether a hadron or resonance can be formed or not.
With these probabilities, the parton and hadron time evolution equations with 
color screening can now be written in the following forms
\be {{\dd f_i^{\mrm{p}}} \over {\dd \t}} \, \Big |_{p_z \t =\mrm{const.}} 
    \; = - \; { {f_i^{\mrm{p}}-f^{\mrm{p}}_{i\; eq}}
            \over {\q_i^{\mrm{p}}} }
    \; =   \; \Big (C_i^{\mrm{p}} - C^{\mrm{p}}_{i\;q_a\bar q_a'} \Big )
            + C'^{\,\mrm{p}}_{i\;q_a\bar q_a'}
            + C'^{\,\mrm{p}}_{i\; g\ra q} + C'^{\,\mrm{p}}_{i\; p\ra h}
            + C'^{\,\mrm{p}}_{i\; h\ra p}    \; ,
\label{eq:cs_p_ev}
\ee
\be {{\dd f_i^{\mrm{h}}} \over {\dd \t}} \, \Big |_{p_z \t =\mrm{const.}} 
    \; = - \; { {f_i^{\mrm{h}}-f^{\mrm{h}}_{i\; eq}}
            \over {\q_i^{\mrm{h}}} }
    \; =   \; C'^{\,\mrm{h}}_{i\;p \ra h} 
            + C'^{\,\mrm{h}}_{i\; h\ra p} \; .
\label{eq:cs_h_ev}
\ee
$C^{\mrm{p}}_{i\;q_a\bar q_a'}$ is the color singlet $q\bar q'$ scattering
term and the primed $C'$'s are almost the same terms as in 
\ederef{eq:new_p_ev}{eq:new_h_ev} above but are now weighed by either $P_<$ 
or $P_>$. The new terms $C_{i\;h\ra p}$ describe the melting of hadrons 
as discussed above. Their forms, further details of the above 
equations and discussions are given in \cite{wong3}.

\section{Color Screening Effects on Hadronization in a Strongly Interacting
Parton Plasma}
\label{sec:cs}

Solving for the distributions from the color screened and unscreened 
equations above, we can compare the effects of the medium on hadronization 
and vice versa \cite{wong3}. In \fref{f:hn&pf}, the $\p$ and $K$ number 
densities $n_h$ against time are plotted on the left figure and the parton 
fugacities $l_g, l_q$ on the right. In the $n_h$ plot, from top to bottom 
three pairs of results, color screened (solid) and unscreened (dashed), 
are shown for the $\p^\pm$, $\p^0$ and $K^{\pm,0}$, respectively. Clearly 
there is a slow start and a delay for forming hadrons in a properly color 
screened plasma because of the struggle between confinement and color screening. 
Consequently, the maximum densities are lower because of the expansion.
In the second plot on the right, one effect of hadronization on the medium
is shown. The top (bottom) three curves are the $l_g$ ($l_q$) results. 
The long dashed, dotted and solid curves are for the case with no 
hadronization, with hadronization but no screening and with both 
included, respectively. So parton equilibration is seen to be 
seriously disrupted by confinement. Unless equilibration is extremely
fast and could finish well before the phase transition which is unlikely, 
fully equilibrated quark-gluon plasma should not be expected. 
In the two cases with hadronization, the disruption to the partons is again
delayed when screening is included. So any model without screening will not
get the hadronization time scale correct. Other effects such as lower hadron
densities caused by the delay will affect estimates on the background 
contributions from the hadron gas to the proposed signatures for the 
quark-gluon plasma. In view of the imminent operation of RHIC a proper 
incorporation of medium effects in numerical models with parton-hadron 
transition is urgently needed. 

\bfi
\centerline{
\epsfig{figure=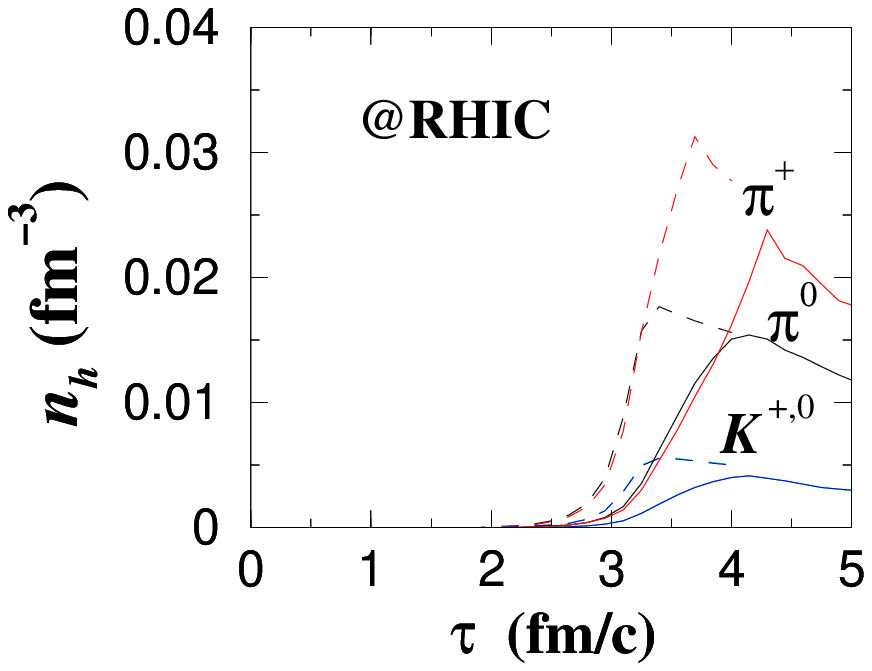,width=2.60in} \hspace{0.70cm}
\epsfig{figure=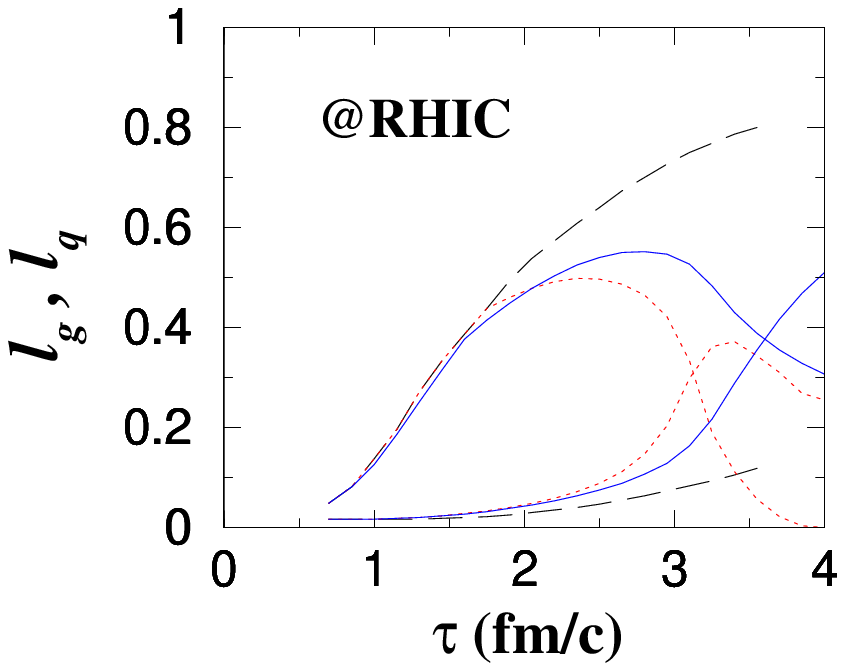,width=2.60in}
}
\vspace{-0.8cm}
\caption{The time developments of the hadron number densities and parton fugacities
in a parton plasma going through hadronization with and without color screening.}
\label{f:hn&pf}
\efi

\end{document}